\documentclass[prc,twocolumn]{revtex4-1}

\usepackage{graphicx}
\usepackage{dcolumn}
\usepackage{bm}
\usepackage{slashed}
\usepackage{amsmath,graphicx}
\usepackage[colorlinks=true,linktocpage=true,linkcolor=blue,citecolor=blue]{hyperref}
\usepackage{float}
\usepackage{nicefrac}
\usepackage[normalem]{ulem}
\usepackage{amsmath}
\usepackage{subfigure}

\begin{document}
 
\def\feq{f_{\rm eq}} 
\def\trel{\tau_{\rm rel}}
\def\t0{{\tau_0}}
\def\tp{{\tau^\prime}}
\def\tb{{\tau^{\prime \prime}}}

\def\chieps{\chi_\varepsilon\!\left(\frac{\mu}{T}\right)}
\def\chiepsinit{\chi_\varepsilon\!\left(\frac{\mu_0}{T_0}\right)}
\def\chiepsp{\chi_\varepsilon\!\left(\frac{\mu^\prime}{T^\prime}\right)}

\def\chiden{\chi_n\!\left(\frac{\mu}{T}\right)}
\def\chideninit{\chi_n\!\left(\frac{\mu_0}{T_0}\right)}
\def\chidenp{\chi_n\!\left(\frac{\mu^\prime}{T^\prime}\right)}

\def\tel{\tau_{\rm el}}
\def\telprime{\tau_{\rm el}^\prime}
\def\telbis{\tau_{\rm el}^{\prime \prime}}
\def\tf{{\tau_{\infty}}}
\def\fst{f_{\rm st}(\mu,T)}
\def\finf{f_\infty}

\def\tin{\tau_{\rm in}}
\def\tinprime{\tau_{\rm in}^\prime}
\def\tinbis{\tau_{\rm in}^{\prime \prime}}

\def\mup{\mu^{\prime}}
\def\Tp{{T^{\prime}}}

\def\pper{p_\perp}
\def\vpper{{\bf p}_\perp}
\def\ppar{p_\parallel}

\def\eps{\varepsilon}
\def\Ppar{P_\parallel}
\def\Pper{P_\perp}

\def\calH{{\cal H}}

\newcommand{\beq}{\begin{eqnarray}}
\newcommand{\eeq}{\end{eqnarray}}

 
\title{Separation of elastic and inelastic processes \\ in the relaxation time approximation for collision integral}

\author{Wojciech Florkowski} 
\email{Wojciech.Florkowski@ifj.edu.pl}
\affiliation{The H. Niewodnicza\'nski Institute of Nuclear Physics, Polish Academy of Sciences, PL-31342 Krak\'ow, Poland}

\author{Radoslaw Ryblewski} 
\email{Radoslaw.Ryblewski@ifj.edu.pl}
\affiliation{The H. Niewodnicza\'nski Institute of Nuclear Physics, Polish Academy of Sciences, PL-31342 Krak\'ow, Poland}

\date{\today}

\begin{abstract}
We introduce a generalised relaxation-time-approximation form of the collision term in the Boltzmann kinetic equation that allows for using different relaxation times for elastic and inelastic collisions.  The efficacy of the proposed framework is demonstrated with the numerical calculations that describe systems with different relations between the two relaxation times and the evolution time of the system. 
\end{abstract}



\maketitle 

\section{Introduction}
\label{sect:intro}

The successful use of perfect and viscous hydrodynamics to explain behaviour of matter produced in relativistic heavy-ion collisions at RHIC and the LHC~\cite{florkowski2010,Heinz:2013th,Gale:2013da} has initiated a broad interest in the formulations of relativistic hydrodynamics and relativistic kinetic theory. In fact, the latter is very often used as an underlying theory for the former~\cite{Jeon:1995zm,Baier:2006um,Romatschke:2009im,Calzetta:2010au,Denicol:2012cn,Denicol:2012es,Denicol:2014loa,Strickland:2014pga,Jeon:2015dfa,Tsumura:2015fxa}. 

Derivation of the equations of relativistic hydrodynamics from the kinetic theory is quite complicated, mainly, because of the complex structure of the collision term which appears in the kinetic theory. Therefore, one often uses a simplifying assumption about the collision term and treats it in the relaxation time approximation (RTA)~\cite{Bhatnagar:1954zz,Anderson:1974,Anderson:1974q,Gavin:1985ph,Czyz:1986mr,Dyrek:1986vv,Jaiswal:2013npa,Jaiswal:2014raa,Nopoush:2014pfa,Jaiswal:2014isa,Florkowski:2015lra}. Within this approximation, the collision term has the structure~\cite{Anderson:1974}
\beq
C = p\cdot u \,\frac{\feq - f}{\trel},
\label{standard}
\eeq
where $f=f(x,p)$ is the parton phase-space distribution function depending on the parton space-time coordinates $x$ and momentum $p$, $u=u(x)$ is the four-vector describing hydrodynamic flow at the space-time point $x$, $\feq$ is the reference equilibrium distribution function, and $\trel$ is the relaxation time. 

The reference equilibrium distribution function $\feq$ has the standard form (Bose-Einstein, Fermi-Dirac or Boltzmann) which depends on the temperature $T$ and chemical potential $\mu$. These two parameters should be treated as effective ones, i.e., they are usually fixed by the two conditions specifying that $\feq$ and $f$ give locally the same energy density and particle number density (Landau matching conditions). Only, if the system is close to the genuine local equilibrium, $T$ and $\mu$ can be interpreted as real temperature and chemical potential. 

In this work we use the Landau definition of the hydrodynamic flow (Landau energy frame). The four-vector $u$ is defined as the direction of the energy flux obtained with the system's energy momentum tensor $T^{\mu\nu}$, namely
\beq
T^\mu_{\,\,\,\nu} u^\nu = \varepsilon u^\mu.
\label{LandauFrame}
\eeq
Here $\varepsilon$ is the local energy density. The local rest frame (LRF) at the space-time point $x$ is defined by the condition $u^\mu(x)=(1,0,0,0)$,  The equilibrium distribution $\feq$ depends on the particle energy measured in the local rest frame, i.e., on $p\cdot u$.

The relaxation time appearing in (\ref{standard}) sets the overall timescale for the approach of the system towards (local) equilibrium. In many applications it is treated as a constant. In more realistic approaches one should include the dependence of $\trel$ on the local properties of matter (typically characterised by $T$ and $\mu$) or on the momentum of interacting particles~\cite{Florkowski:2015dmm,Jaiswal:2016sfw}. As the use of the medium dependent relaxation time is rather straightforward, the inclusion of the momentum dependence is much more complicated. Another aspect of using RTA is that the relaxation time should depend on the type of interaction and is different for elastic and inelastic collisions~\cite{Blaizot:2016bsg,Huang:2013lia}.

In this paper we show that the formula (\ref{standard}) can be naturally generalised to the situation where the elastic and inelastic processes are characterised by two different timescales, $\tel$ and $\tin$, respectively. The newly proposed equation is solved in the case of simple one-dimensional expansion to demonstrate that the obtained solutions exhibit behaviour expected in the situations characterised by different relations between the relaxation times $\tel$ and $\tin$.

We note that in high-energy processes the number of particles is not conserved, hence, the chemical potential $\mu$ is not taken into account. It is replaced by the baryon chemical potential $\mu_B$ in the situations where one deals with different particle species that have different baryon numbers.  In the present work we show that it makes sense to introduce the {\it effective chemical potential} in the situations where we have two different timescales for elastic and inelastic processes (even for simple, one component systems). Depending on the mutual relations between $\tel$, $\tin$, and the system's evolution time $\tau$, the effective chemical potential may vanish or gain a non-zero value.

\medskip
The structure of the paper is as follows: In Sec.~\ref{sect:kineq} we define the new form of the RTA kinetic equation that allows for separation of elastic and inelastic processes.  In Sec.~\ref{sect:expansion} we consider one-dimensional boost-invariant expansion and present formal, analytic solutions of the new kinetic equation in this case.  The results of numerical calculations for the one-dimensional expansion are presented in Sec.~\ref{sect:results}. We conclude and discuss possible future applications of our approach in Sec.~\ref{sect:conclusions}.

\medskip
{\it Notation and conventions:} Throught the paper we use natural units with $c = \hbar = k_B = 1$. We also use the notation $p = (E, \vpper, \ppar)$ for the four-momentum of a particle, and $\pper = |\vpper |$.  The invariant integration measure in the momentum space $g d^3p/((2\pi)^3 E) $, with $g$ being the internal degeneracy factor, is denoted shortly as $dP$. The metric tensor has the signature $(+,-,-,-)$. The scalar product of two four-vectors $a$ and $b$ is denoted as $a \cdot b$. For sake of simplicity, we consider the case of massless particles, setting $m=0$.

\section{Generalised RTA kinetic equation}
\label{sect:kineq}

\subsection{Kinetic equation}

In this work we propose to use the kinetic equation 
\beq
 p^\mu \partial_\mu  f(x,p) =  C[ f(x,p)],
\label{kineq}
\eeq
with the collision term  $C$ of the form
\begin{eqnarray}
C[f] =  p \cdot u  \left[
\frac{\feq(\mu,T)-f}{\tel} + \frac{\feq(0,T)-f}{\tin}
\right] .
\label{col-term}
\end{eqnarray}
The quantities $\tel$ and $\tin$ are the relaxation times characterising elastic (particle number conserving) and inelastic (particle number changing) processes, respectively.  In general, $\tel$ and $\tin$ may depend on $x$ and $p$. In this work, we restrict our considerations to the case where  $\tel$ and $\tin$ are both constant.

The equilibrium distribution function $\feq$ may be taken to be a Bose-Einstein, Fermi-Dirac, or Boltzmann distribution. These three cases correspond to the form
\begin{eqnarray}
\feq(\mu,T) =  \left[ \exp\left(\frac{p \cdot u - \mu}{T} \right) - \epsilon  \right]^{-1}
\label{eqdistr}
\end{eqnarray}
with $\epsilon=1, -1$~and~0, respectively.

In our notation we display only thermodynamic arguments of the equilibrium functions to emphasise that the term describing elastic (inelastic) processes has $\feq$ with finite (vanishing) chemical potential. The dependence of $\feq$ on momentum and space-time coordinates (through $\mu(x)$, $T(x)$, and $u(x)$) is not shown.

The idea behind proposing the form (\ref{kineq}) is that it naturally describes the tendency of a system to approach local equilibrium ---  inelastic processes lead to local equilibrium defined by the temperature only, while elastic processes lead to equilibrium described by the temperature and chemical potential.

\subsection{Energy-momentum conservation}

Using Eqs.~(\ref{kineq}),  (\ref{col-term}), and the standard definition of the energy-momentum tensor as the second moment of the distribution function, namely
\beq
T^{\mu\nu} = \int dP\, p^\mu p^\nu f,
\label{Tmunu}
\eeq
we find that the energy is conserved, $\partial_\mu T^{\mu\nu} = 0$, if the following condition is satisfied
\beq
 \eps = \frac{3 g T^4}{\pi^2}  \left(  \frac{\chieps \tin + \chi_\varepsilon(0)\,\tel}{\tin + \tel}  \right).
 \label{LMe}
\eeq
Here $\eps$ is the non-equilibrium energy density obtained as the projection $\eps = u_\mu u_\nu T^{\mu\nu}$, while the function $\chi_\varepsilon$ is defined as~\cite{florkowski2010}
\beq
\chieps =
\begin{cases}
\hbox{Li}_4\left(e^\frac{\mu}{T} \right)   & \hbox{for Bose-Einstein,} \\
-\hbox{Li}_4\left(-e^\frac{\mu}{T} \right) & \hbox{for Fermi-Dirac,}\\
e^\frac{\mu}{T}  & \hbox{for Boltzmann.} 
\end{cases}
\label{chieps}
\eeq
The function $\hbox{Li}_s(z)$ is the polylogarithm function defined by the series
\beq
\hbox{Li}_s(z) = \sum_{k=1}^\infty \frac{z^k}{k^s}.
\eeq
Equation (\ref{LMe}) can be treated as a generalised Landau matching condition that can be used to determine  $T$ and~$\mu$. 

\subsection{Particle number conservation in elastic collisions}

The second necessary condition may be obtained from the analysis of the particle number current
\beq
N^{\mu} = \int dP\, p^\mu  f.
\eeq
Using again Eqs.~(\ref{kineq}) and  (\ref{col-term}) we find
\beq
\partial_\mu N^\mu = \left( \partial_\mu N^\mu \right)_{\rm el}  + \left( \partial_\mu N^\mu \right)_{\rm in}, 
\label{divN}
\eeq
where
\beq
\left( \partial_\mu N^\mu \right)_{\rm el} = \frac{1}{\tel} \left( \frac{g T^3 }{\pi^2} \,\chiden - n  \right)
\label{divNel}
\eeq
and
\beq
\left( \partial_\mu N^\mu \right)_{\rm in} = \frac{1}{\tin} \left( \frac{g T^3}{\pi^2} \,\chi_n(0) - n  \right).
\label{divNin}
\eeq
Similarly to (\ref{chieps}) we have~\cite{florkowski2010}

\beq
\chiden =
\begin{cases}
\hbox{Li}_3\left(e^\frac{\mu}{T} \right)   & \hbox{for Bose-Einstein,} \\
-\hbox{Li}_3\left(-e^\frac{\mu}{T} \right) & \hbox{for Fermi-Dirac,}\\
e^\frac{\mu}{T}  & \hbox{for Boltzmann.} 
\end{cases}
\eeq

The right-hand side of~Eq.~(\ref{divN}) describes the production of particles due to the presence of elastic and inelastic processes.
Although we cannot assume that $\partial_\mu N^\mu=0$, we should require that elastic processes do not contribute to the particle production, hence we use the condition
\beq
\left( \partial_\mu N^\mu \right)_{\rm el} =0,
\eeq
which leads to 
\beq
n =  \frac{g T^3 }{\pi^2} \chiden.
\label{LMd}
\eeq 

Equation (\ref{LMd}) is the second Landau matching condition, which together with Eq.~(\ref{LMe}) allows us to determine both $T$ and $\mu$. We note that $\eps$ and $n$ appearing on the left-hand sides of Eqs.~(\ref{LMe}) and (\ref{LMd}) are non equilibrium quantities, so $T$ and $\mu$ are an alternative way to define $\eps$ and $n$. 

\medskip
At this point we want to stress that Eqs.~(\ref{kineq}), (\ref{LMe}), and (\ref{LMd}) represent a complete system of three equations that define our framework. In the remaining part of this section, we discuss its main properties.

\subsection{Late time asymptotics}

If the evolution time of the system $\tau$ is much larger than the two relaxation times, 
\beq
\tel, \tin \ll \tau,
\label{caseI}
\eeq
one expects that the system approaches local equilibrium state where the right-hand side of the kinetic equation (\ref{kineq}) vanishes.   In this case  the distribution function has the form
\beq
\fst = \frac{\,\feq(\mu,T)\tin +\,\feq(0,T)  \tel}{\tin+\tel} .
\label{finaldistr}
\eeq
Using (\ref{finaldistr}) one may check that the Landau matching for the energy density (\ref{LMe}) is automatically fulfilled, but from the Landau matching for the particle number density (\ref{LMd}) one gets
\beq
\frac{\chiden \tin +  \chi_n(0)\tel}{\tin+\tel}  = \chiden.
\label{2LM}
\eeq
Equation~(\ref{2LM}) implies directly that the asymptotic equilibrium state is achieved only with $\mu=0$. This property reflects the fact that inelastic processes are present in the system, and for sufficiently large evolution times, see~(\ref{caseI}), they force $\mu$ to vanish. 

Similar situation happens if the evolution time is much larger than $\tin$ but shorter than $\tel$,
\beq
\tin \ll \tau \ll \tel.
\label{caseII}
\eeq
In this case the system is very close to the equilibrium state determined by particle changing processes (note that taking a formal limit $\tin \to 0$ in (\ref{2LM}) one immediately finds $\mu=0$).

\subsection{Transient equilibrium with conserved \\ number of particles}

Interesting situation takes place in the case
\beq
\tel \ll \tau \ll \tin.
\label{caseIII}
\eeq
This corresponds to the situation where the system is dominated by elastic collisions. Taking the limit $\tel \to 0$ in (\ref{2LM}) one finds that this equation is always fulfilled.

For the evolution time $\tau$ satisfying Eq.~(\ref{caseIII}) the system behaves almost like a perfect fluid. The parameters $T$ and $\mu$ are determined by the hydrodynamic equations. In this case the number of particles and entropy are both conserved, hence we can write $\partial_\mu N^\mu = 0$ and $\partial_\mu S^\mu = 0$, where $N^\mu = n u^\mu$ and $S^\mu = s u^\mu$. For massless particles the entropy density $s$ is connected with the particle density through the relation
\beq
s = 4 n \left[ \frac{\chieps}{\chiden} - \frac{\mu}{4 T}\right].
\eeq
Consequently, the conservation laws for the particle number and entropy lead to the condition
\beq
u^\mu \partial_\mu \left(\frac{\mu}{T}\right)=0.
\label{muThydro}
\eeq
Thus, for the system dominated by elastic collisions, see~ (\ref{caseIII}), we expect that the ratio $\mu/T$ is constant along the world lines of fluid elements.

\section{Boost-invariant expansion}
\label{sect:expansion}

\subsection{Implementation of the symmetry constraints}
\label{sect:sym}

In the case of one-dimensional boost-invariant expansion, all scalar functions of time and space depend only on the proper time $\tau = \sqrt{t^2-z^2}$, and the hydrodynamic flow has the form $u^\mu= \left({t}/{\tau},0,0,{z}/{\tau}\right)$ \cite{Bjorken:1982qr}. In addition, the phase-space densities behave like scalars under Lorentz transformations, hence, $f(x,p)$ may depend only on: $\tau$, $w$ and $\vpper$, where $w =  t \ppar - z E$~ \cite{Bialas:1984wv,Bialas:1987en}. Using $w$ and $\vpper$ we define another convenient variable
\begin{equation}
v(\tau,w,\pper) = Et - \ppar z = 
\sqrt{w^2+\vpper^{\,\,2} \tau^2} \, .  
\label{v}
\end{equation}
Thus, the momentum integration measure $dP$ in phase-space may be written as
\begin{equation}
dP =  \frac{g\,d\ppar\,d^2\pper}{(2\pi)^3 E} =  \frac{g\,dw\,d^2\pper }{(2\pi)^3 v}. 
\label{dP}
\end{equation}
With the help of such boost-invariant variables one finds 
\beq
\partial_\mu u^\mu = \frac{1}{\tau}, \quad u^\mu \partial_\mu  = \frac{d}{d\tau}, \quad 
p \cdot u = \frac{v}{\tau} 
\label{notation}
\eeq
and
\begin{eqnarray}
p^\mu \partial_\mu f = 
\frac{v}{\tau} \frac{\partial f}{\partial \tau}\, .
\label{binvterms}
\end{eqnarray}
Using Eqs.~(\ref{notation})  and (\ref{binvterms}) in Eq.~(\ref{kineq}) one finds a simple form of the original kinetic equation, namely
\begin{eqnarray}
\frac{\partial f}{\partial \tau}  &=& 
\frac{\feq(\mu,T)-f}{\tel} + \frac{\feq(0,T)-f}{\tin} \, ,
\label{kineqtau}
\end{eqnarray} 
where the equilibrium distribution function may be written as
\begin{eqnarray}
\feq(\tau,w,\pper) =  \left\{ \exp\left[ \frac{ \sqrt{w^2/\tau^2+\pper^2} -\mu(\tau)}{T(\tau)}  \right]\!-\!\epsilon \right\}^{-1}.
\nonumber \\ \label{eqditau}
\end{eqnarray}
In the following we assume that 
$f(\tau,w,\vpper)$ 
is an even function of $w$ and depends only on the magnitude of the transverse momentum $\pper =|\vpper|$, and $
f(\tau,w,\pper) = f(\tau,-w,\pper)$.

The formal analytic solution of the kinetic equation (\ref{kineqtau}) is~\cite{Baym:1984np,Baym:1985tna,Florkowski:2013lza,Florkowski:2013lya}
\beq
f (\tau,w,\pper) &=& D(\tau,\tau_0) f_0(w,\pper) \label{sol} \\
& + & \int_\t0^\tau d\tp D(\tau,\tp)
\left[ \frac{\feq(\mup,\Tp)}{\tel} + \frac{\feq(0,\Tp)}{\tin} \right],
\nonumber
\eeq
where the damping function $D(\tau_2,\tau_1)$ is defined as
\beq
D(\tau_2,\tau_1) = \exp\left[ - \frac{(\tau_2-\tau_1) (\tel + \tin)}{\tel \tin} \right].\label{damp}
\eeq
Here we use the notation $\mup = \mu(\tp)$, $\Tp = T(\tp)$. 

\subsection{Energy-momentum conservation and particle number conservation in elastic processes}
\label{sect:enmomcon}

Using the symmetry properties discussed in the previous section, we may rewrite (\ref{Tmunu}) in the form 
\cite{Florkowski:2010cf,Martinez:2012tu}
\begin{equation}
T^{\mu\nu} = (\eps + \Pper) u^\mu u^\nu - \Pper g^{\mu\nu} + (\Ppar-\Pper) z^\mu z^\nu \, , 
\label{Tmunu2}
\end{equation}
where
\beq
\eps(\tau) &=& \frac{1}{\tau^2}\,
\int dP \, v^2\,  f(\tau,w,\pper)  \label{endens}
\eeq
is the energy density, 
\beq
\Ppar(\tau) &=& \frac{1}{\tau^2}\,
\int dP \, w^2\,  f(\tau,w,\pper) \label{longpres}
\eeq
is the longitudinal pressure, and 
\beq
\Pper(\tau) &=& \frac{1}{2}\,
\int dP \, \pper^2\, f(\tau,w,\pper)      \label{transvpres}
\eeq
is the transverse pressure. In (\ref{Tmunu2}) $z^\mu = \left({z}/{\tau},0,0,{t}/{\tau}\right)$ is the four-vector orthogonal to $u^\mu$ which defines the longitudinal direction.

The energy-momentum conservation law is expressed by the vanishing divergence of the energy-momentum tensor $\partial _\mu T^{\mu \nu }(x)=0$. In our case these four equations are reduced to the condition
\begin{eqnarray}
\frac{d\eps}{d\tau}=
- \frac{\eps+\Ppar}{\tau} \, . \label{enmomcon2} \\
\nonumber 
\end{eqnarray}
Calculating the proper time derivative of the  energy density defined by Eq.~(\ref{endens}) and using Eqs.~(\ref{kineqtau}), (\ref{longpres}), and (\ref{enmomcon2}) we find the formula
\beq
g \int \frac{dw d^2\pper}{(2\pi)^3} \,\frac{v}{\tau^2}\, f   
= \frac{3 g T^4}{\pi^2}  \left( \frac{\chieps \tin + \chi_\varepsilon(0) \tel}{\tin+\tel}  \right), \nonumber \\
\label{LMenergy1}
\eeq
where the distribution function $f$ should be taken as the solution (\ref{sol}). We note that (\ref{LMenergy1}) is nothing else but a special case of the Landau matching condition for energy~(\ref{LMe}).

Similarly to the energy conservation, we can discuss the particle number conservation by elastic processes. For our system, $N^\mu = n u^\mu$, and the particle density $n$ is defined by the integral
\beq
n(\tau) = \frac{1}{\tau} \int dP\, v\, f(\tau, w, \pper) .
\label{density} 
\eeq
From the Landau matching condition for particle number density, see Eq.~(\ref{LMd}), we obtain 
\beq
n = g   \int \frac{dw d^2\pper}{(2\pi)^3 \tau}  f  = \frac{g T^3 }{\pi^2} \chiden.
\label{LMdensity1}
\eeq
Here again the function $f$ should be taken as the formal solution (\ref{sol}).

\begin{widetext}
The integrals over momenta in Eqs.~(\ref{LMenergy1}) and (\ref{LMdensity1}) can be done analytically, which yields
\beq
 \frac{ \chieps \tin +  \chi_\varepsilon(0) \tel}{\tin + \tel} \,\,T^4 
 = \frac{1}{2} D(\tau,\tau_0) \chiepsinit \calH\left(\frac{\t0}{\tau}\right) T_0^4 
 +  \frac{1}{2} \int_\t0^\tau \!\!d\tp  D(\tau,\tp) \calH\left(\frac{\tp}{\tau} \right) 
 \Tp^4 \left(   \frac{\chiepsp}{\tel} + \frac{\chi_\varepsilon(0)}{\tin} \right)
 \label{eqforT}
 \eeq
and
 \beq
\chiden  \,T^3 =  D(\tau,\tau_0) \chideninit \frac{\t0 }{\tau} \,T_0^3
 +  \int_\t0^\tau  \!\!d\tp D(\tau,\tp) \frac{\tp}{\tau} \, \Tp^3
 \left(  \frac{\chidenp}{\tel} + \frac{\chi_n(0)}{\tin} \right).
 \label{eqformu} \\
 \nonumber
 \eeq
 \end{widetext}
Here we have assumed that the initial distribution function at the time $\tau = \t0$ is also an equilibrium distribution. 

\begin{figure}[t]
\begin{center}
\includegraphics[angle=0,width=0.45\textwidth]{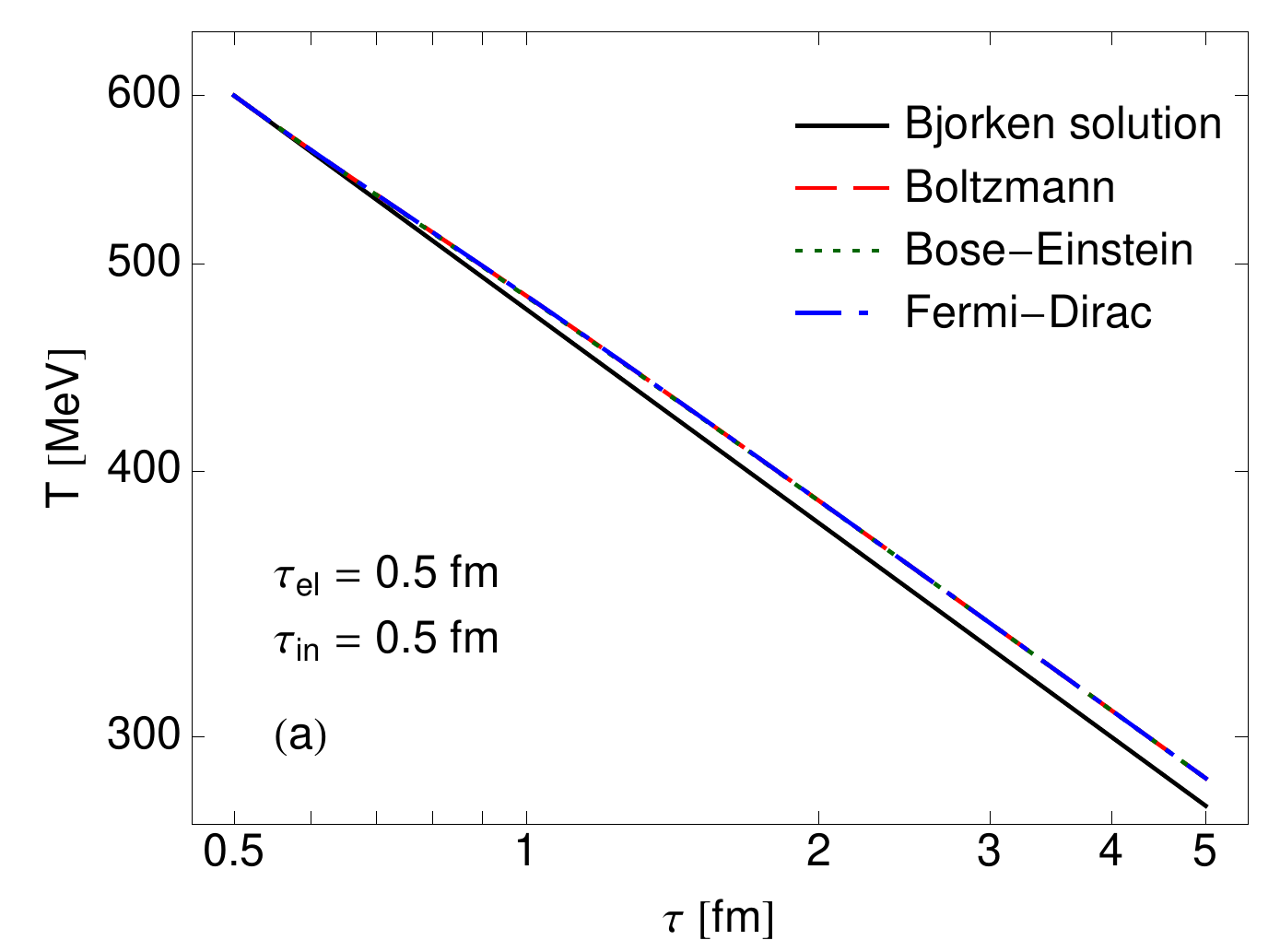} \\
\includegraphics[angle=0,width=0.45\textwidth]{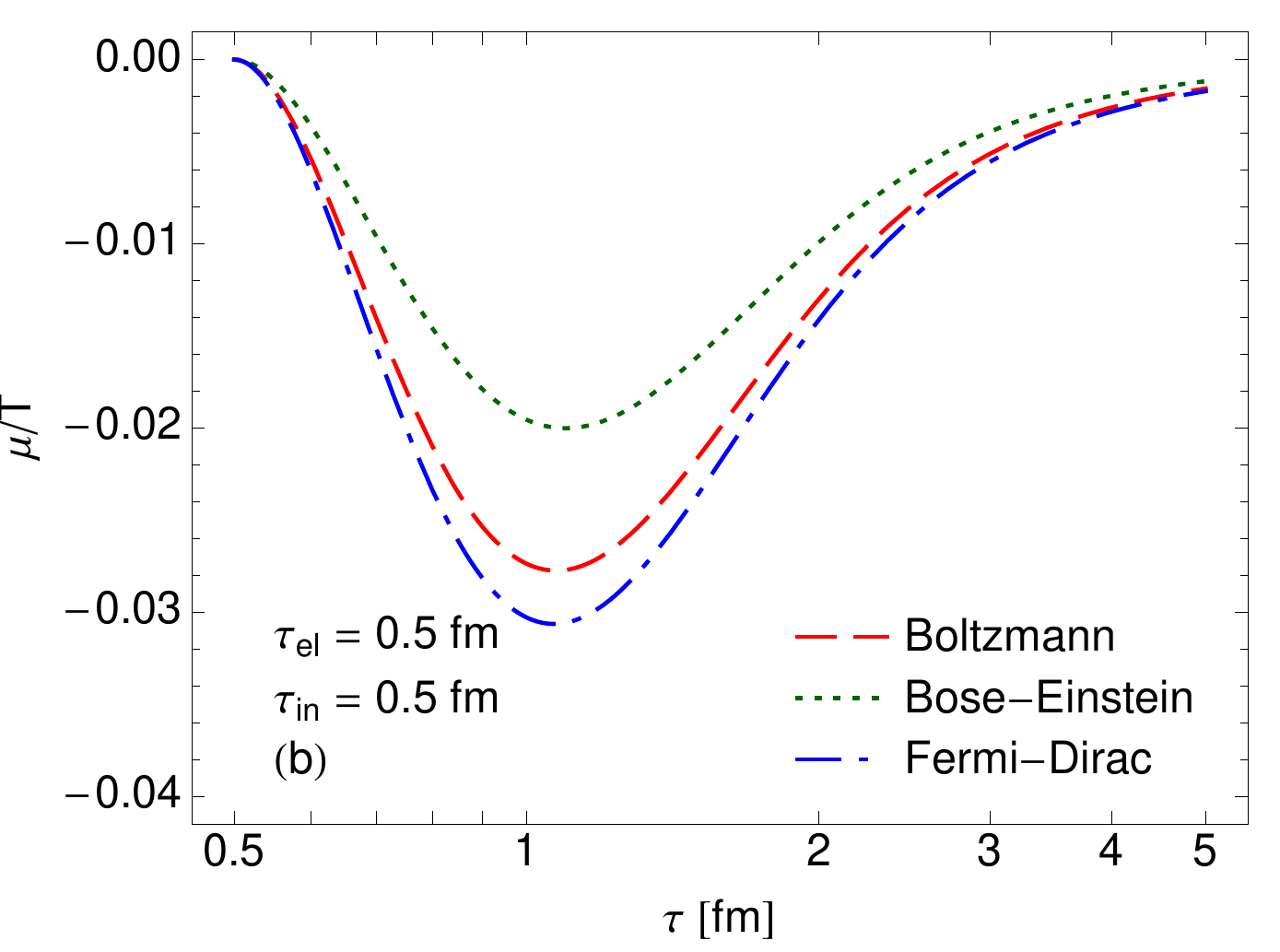}
\end{center}
\caption{(Color online) Time evolution of the effective temperature (a) and effective chemical potential divided by effective temperature (b) determined by Eqs. (3), (7), and (16) with $\tin=\tel=$~0.5~fm. The red dashed, blue dotted-dashed, and green dotted lines describe the results for Boltzmann, Fermi-Dirac, and Bose-Einstein statistics, respectively.  The solid line in the upper panel shows the Bjorken solution for the temperature profile, $T = T_0 (\tau_0/\tau)^{1/3}$. Here $\t0=$~0.5~fm, $T_0=$~600~MeV, and $\mu_0=$~0.}
\label{fig:inter}
\end{figure}
The function $\calH(x)$ appearing in Eq.~(\ref{eqforT}) is defined by the integral~\cite{Florkowski:2013lya}
\beq
\calH(x) = x \int_0^\pi d\phi \, \sin\phi \,\sqrt{x^2 \cos^2\phi + \sin^2\phi }.
\eeq
Equations (\ref{eqforT}) and (\ref{eqformu}) are two integral equations for two functions of the proper time: $T(\tau)$ and $\mu(\tau)$. They can be solved by the iterative method~\cite{Banerjee:1989by}: One substitutes auxiliary functions $T(\tau)$ and $\mu(\tau)$ into the right-hand sides of (\ref{eqforT}) and (\ref{eqformu}) to obtain new time profiles $T(\tau)$ and $\mu(\tau)$ from the left-hand sides, which are again inserted  into the right-hand sides. Repeating this procedure, we approach the stable results which do not change after iteration and represent the solutions of (\ref{eqforT}) and (\ref{eqformu}).

To check our numerical results for $T(\tau)$ and $\mu(\tau)$ we calculate in addition the longitudinal pressure and check if Eq.~(\ref{enmomcon2}) is satisfied. Using (\ref{sol}) in (\ref{longpres}) we find
\beq
&&\Ppar = \frac{3 g }{2 \pi^2} \left[ D(\tau,\tau_0) \chiepsinit \calH_\parallel\left(\frac{\t0}{\tau}\right) T_0^4 
\vphantom{ \frac{\chiepsp}{\tel} }
\right.  \label{PL} \\
&&  + \left.  \int_\t0^\tau \!\!d\tp D(\tau,\tp) \calH_\parallel\left(\frac{\tp}{\tau} \right) 
 \Tp^4 \left(   \frac{\chiepsp}{\tel} + \frac{\chi_\varepsilon(0)}{\tin} \right) \right],
\nonumber
 \eeq
where the function $\calH_\parallel(x)$ is defined by the integral~\cite{Florkowski:2013lya}
\beq
\calH_\parallel(x) = x^3 \int_0^\pi d\phi \, \frac{\sin\phi \cos^2\phi}{\sqrt{x^2 \cos^2\phi + \sin^2\phi }}.
\eeq
We note that analytic expressions for $\calH(x)$ and $\calH_\parallel(x)$ are given in~\cite{Florkowski:2013lya}.
 
\section{Numerical results}
\label{sect:results}

%
\begin{figure}[t]
\begin{center}
\includegraphics[angle=0,width=0.45\textwidth]{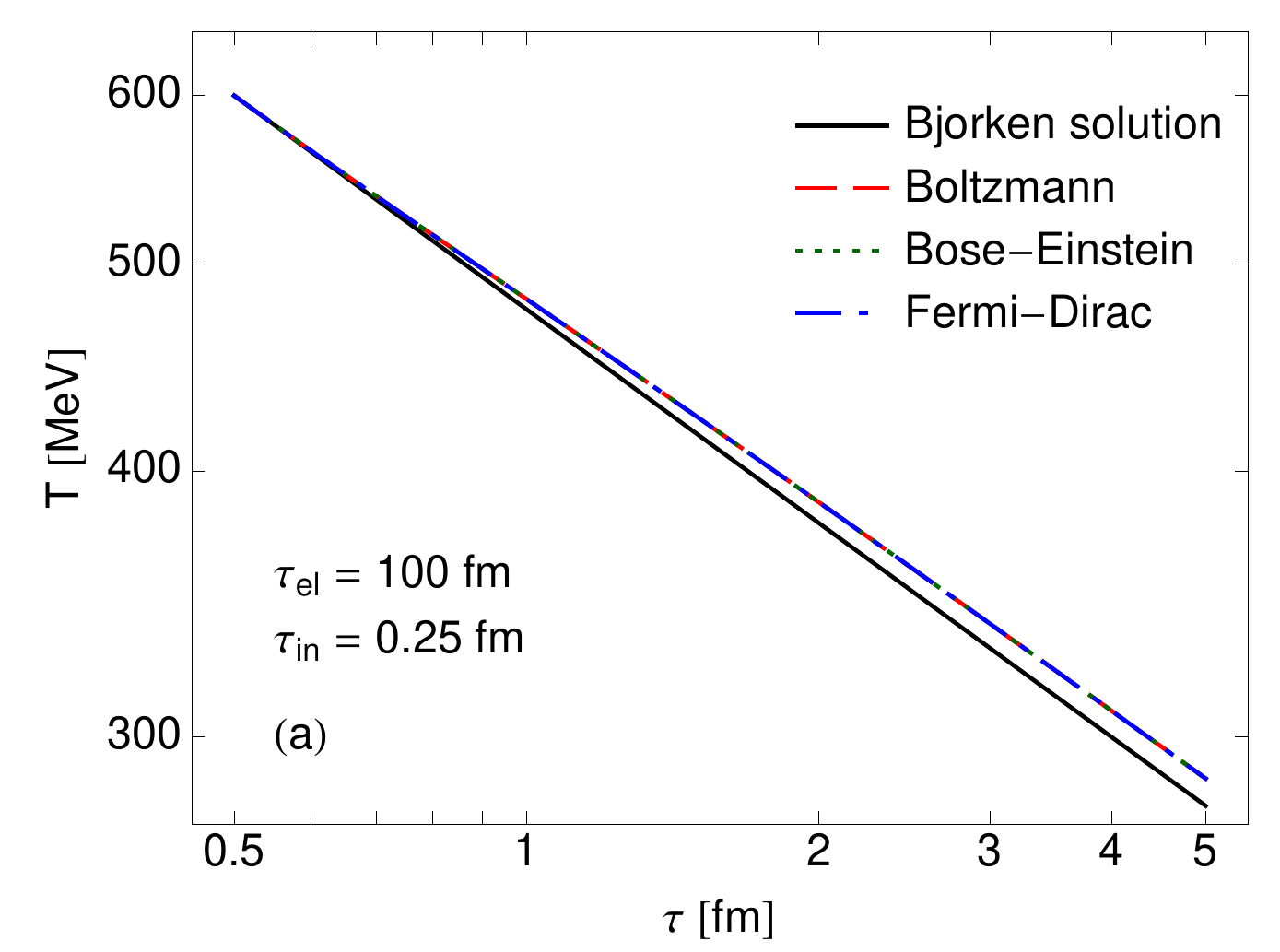} \\
\includegraphics[angle=0,width=0.45\textwidth]{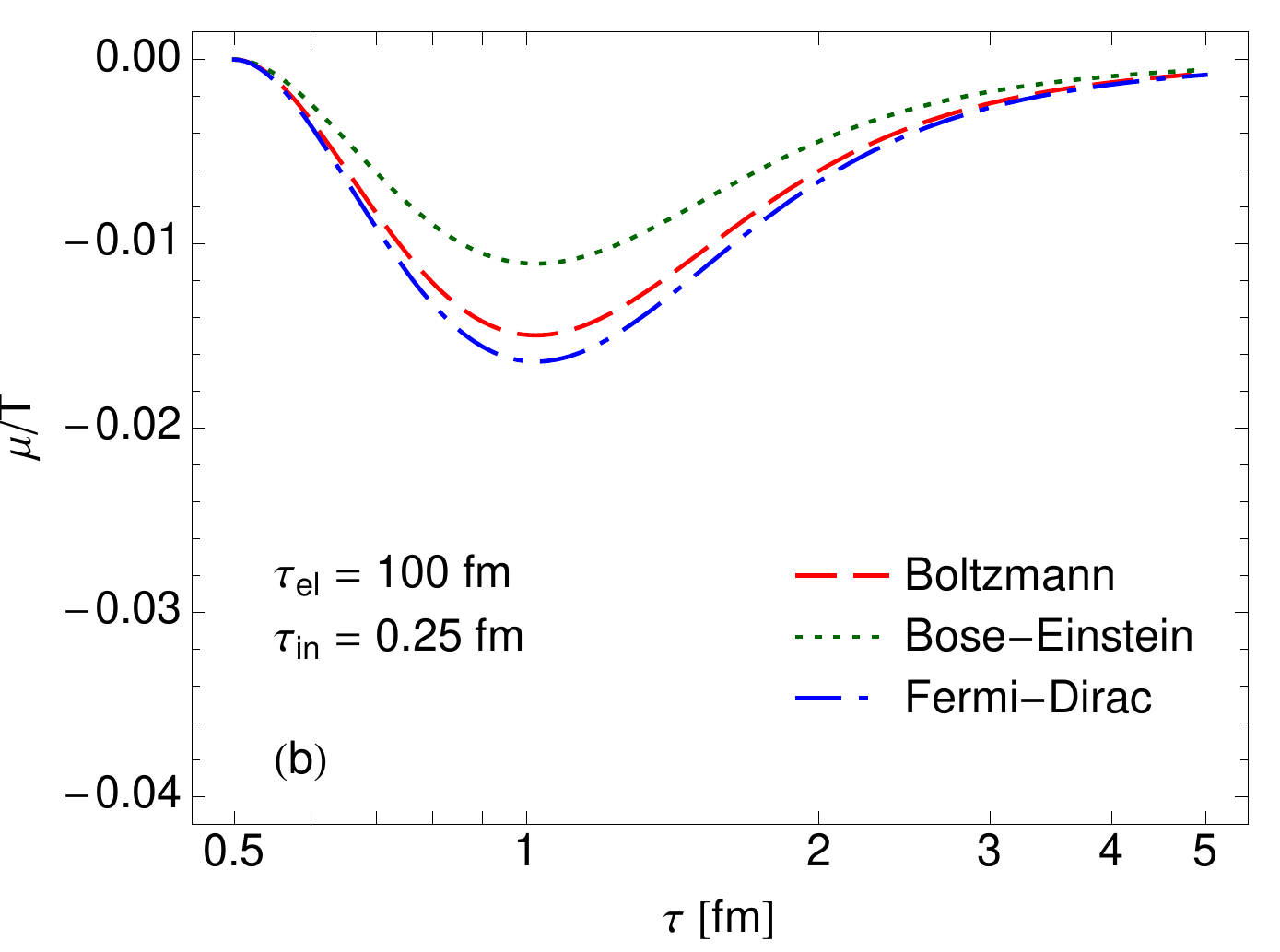}
\end{center}
\caption{(Color online) Same as Fig.~\ref{fig:inter} but with $\tin=$~0.25~fm and $\tel=$~100~fm. }
\label{fig:inel}
\end{figure}
\begin{figure}[t!]
\begin{center}
\includegraphics[angle=0,width=0.45\textwidth]{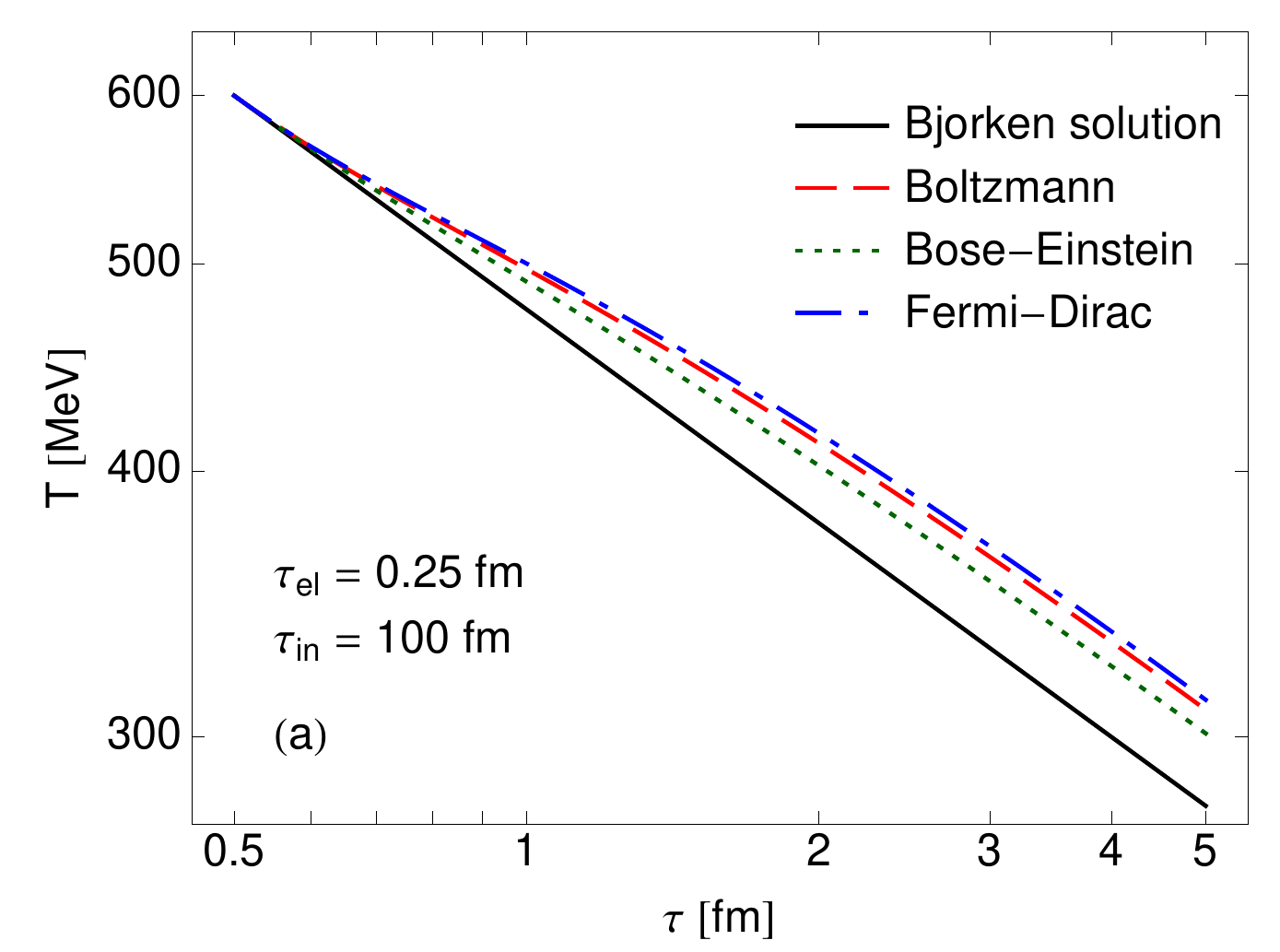} \\
\includegraphics[angle=0,width=0.45\textwidth]{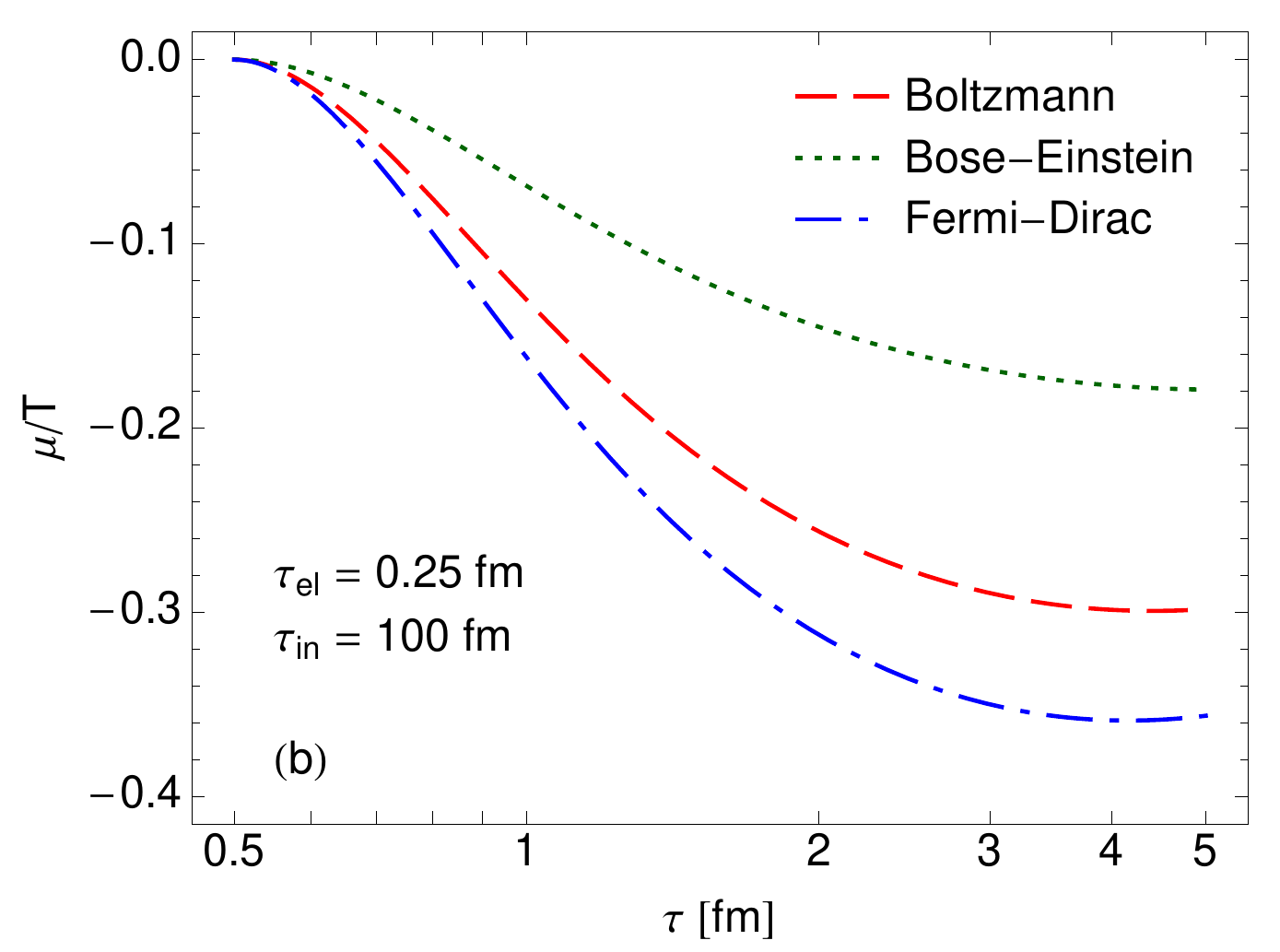}
\end{center}
\caption{(Color online) Same as Fig.~\ref{fig:inter} but with $\tel=$~0.25~fm and $\tin=$~100~fm. }
\label{fig:el}
\end{figure}

In this Section we present the results of numerical calculations based on Eqs.~(\ref{kineq}), (\ref{LMe}), and (\ref{LMd}). Our aim is to illustrate different scenarios discussed earlier in Sec.~\ref{sect:kineq}. Our initial conditions are set at the initial proper time $\tau_0=$~0.5~fm. They correspond to the equilibrium distribution of the form (\ref{eqditau}) with initial temperature $T_0=$~600~MeV and initial chemical potential $\mu_0=$~0. We continue the time evolution of the system till $\tau=$~5~fm.

In Fig.~\ref{fig:inter} we show our results obtained for the case $\tin=\tel=$~0.5~fm. The red dashed, blue dotted-dashed, and green dotted lines describe the results obtained for Boltzmann, Fermi-Dirac, and Bose-Einstein statistics, respectively.  The solid line in the upper panel shows, for the reference,  the Bjorken solution for the temperature profile, $T = T_0 (\tau_0/\tau)^{1/3}$.  We observe that the non-equilibrium behaviour at the beginning of the evolution makes the temperature higher than that found in the Bjorken scenario, while the chemical potential (scaled by the temperature) becomes first negative and later approaches zero. The results for the effective temperature are practically independent of the quantum statistics assumed in the calculations. On the other hand, the results for the chemical potential vary with statistics --- the largest change of the $\mu/T$ ratio is found for the Fermi-Dirac statistics, while the smallest change is found for the Bose-Einstein case. The overall change of the $\mu/T$ ratio is rather small ($\mu/T$ at the minimum reaches -0.03). As expected, due to the presence of inelastic processes, the effective chemical potential tends to zero for large evolution times.

Similar situation to that shown in Fig.~\ref{fig:inter} takes place if $\tin=$~0.25~fm and $\tel=$~100~fm, see Fig.~\ref{fig:inel}. In this case the inelastic processes play a dominant role. The temperature profiles found for different statistics are practically the same as those found in the case  $\tin=\tel=$~0.5~fm. The ratio $\mu/T$ becomes again negative at the beginning of the evolution, but the effect is smaller than that found in the case $\tin=\tel=$~0.5~fm (due to a smaller value of the inelastic collision time, which keeps the system always close to chemical equilibrium with $\mu=0$).

An interesting situation takes place in the case $\tel=$~0.25~fm and $\tin=$~100~fm, see Fig.~\ref{fig:el}. For the evolution times of about a  few fermis the system dynamics is dominated by elastic processes. We observe much larger increase of the effective temperature (compared to the Bjorken scenario) accompanied with a significant decrease of the $\mu/T$ ratio towards negative values (the value at the minimum reaches -0.36). The results do depend on the statistics of particles; the strongest effects are for the Fermi-Dirac case, and the smallest are for the Bose-Einstein case. For $\tau >$~3~fm, the ratio $\mu/T$ is approximately flat, which reflects local equilibrium state described by the hydrodynamic equation~(\ref{muThydro}).

It is also interesting to realise that in all studied cases the initial non-equilibrium dynamics leads to a relative increase of $T$ (compared to Bjorken scaling) and decrease of the ratio $\mu/T$. This seems to be undesirable situation in the context of possible creation of the gluon condensate in heavy-ion collisions~\cite{Blaizot:2011xf,Blaizot:2013lga,Blaizot:2014jna,Xu:2014ega,Blaizot:2015wga,Blaizot:2015xga}, which may lead in turn to pion condensate \cite{Begun:2013nga,Begun:2015ifa}. A signal for the creation of the condensate would be a growth of the chemical potential and reaching the critical value, which is zero in our case (we deal with massless particles). Although we start with the critical value, we find no numerical evidence that $\mu/T$ grows with time. Contrary, this ratio first decreases and only later approaches slowly zero from below. We note that the decrease of $\mu/T$ is the largest in the case where the system is dominated by elastic collisions --- such situation would naively favour the creation of the condensate. Certainly, to complete the picture of condensation further processes and effects should be included such as particle production and momentum dependent relaxation times. This would be an interesting development of the framework introduced in this work.

\section{Conclusions and outlook}
\label{sect:conclusions}

In this paper we have introduced a modified form of the relaxation-time-approximation for the collision term in the Boltzmann equation that allows for separation of elastic and inelastic collisions. We have showed how it can be supplemented consistently by Landau matching conditions for energy and particle number density. 

The proposed scheme offers multiple applications in the situations where one wants to study consequences of having two different relaxation times for elastic and inelastic collisions, but one does not want to invoke the whole machinery of the kinetic theory with complicated collision integrals. The list of possible extra effects that may be taken into account includes: finite masses of particles, 
mixtures, color mean fields, momentum dependent relaxation times, and source terms. The new form of the collision term may be used to derive new equations of dissipative and anisotropic hydrodynamics. Moreover, with the inclusion of source terms descrybing particle production, it may be possible to study conditions allowing for Bose-Einstein condensation.

\medskip
{\bf Acknowledgments:} W.~F. and R.~R. were supported in part by Polish National Science Center Grants No.
DEC-2012/06/A/ST2/00390 and DEC-2012/07/D/ST2/02125, respectively.

\end{document}